# Critical Transit Infrastructure in Smart Cities and Urban Air Quality: A Multi-City Seasonal Comparison of Ridership and $PM_{2.5}$

Sean Elliott
Department of Computer Science
University of Nevada, Las Vegas
Las Vegas, USA
ellios3@unlv.nevada.edu

Sohini Roy
Department of Computer Science
University of Nevada, Las Vegas
Las Vegas, USA
sohini.roy@unlv.edu

***Abstract*—Public transit is a critical component of urban mobility and equity, yet mobility–air-quality linkages are rarely operationalized in reproducible smart-city analytics workflows. This study develops a transparent, multi-source monitoring dataset that integrates agency-reported transit ridership with ambient fine particulate matter $PM_{2.5}$ from the U.S. EPA Air Quality System (AQS) for four U.S. metropolitan areas—New York City, Chicago, Las Vegas, and Phoenix, using two seasonal snapshots (March and October 2024). We harmonize heterogeneous ridership feeds (daily and stop-level) to monthly system totals and pair them with monthly mean $PM_{2.5}$, reporting both absolute and per-capita metrics to enable cross-city comparability. Results show pronounced structural differences in transit scale and intensity, with consistent seasonal shifts in both ridership and $PM_{2.5}$ that vary by urban context. A set of lightweight regression specifications is used as a descriptive sensitivity analysis, indicating that apparent mobility-$PM_{2.5}$ relationships are not uniform across cities or seasons and are strongly shaped by baseline city effects. Overall, the paper positions integrated mobility and environment monitoring as a practical smart-city capability, offering a scalable framework for tracking infrastructure utilization alongside exposure-relevant air-quality indicators to support sustainable communities and public-health-aware urban resilience.**

## I. INTRODUCTION

Ambient fine particulate matter ( $PM_{2.5}$ ) is widely recognized as a key environmental determinant of population health in urban areas. Rather than framing $PM_{2.5}$ solely as a regulatory compliance metric, recent research has emphasized its direct implications for longevity and quality of life. For example, the authors in [1] estimate that global $PM_{2.5}$ exposure reduces average life expectancy by approximately one year, with even larger losses observed in highly polluted regions. By expressing $PM_{2.5}$ exposure in terms of life-expectancy loss, their work highlights that even moderate cross-city or seasonal differences in $PM_{2.5}$ can have meaningful population-level consequences. Similar themes appear in city-level assessments, like [2], which estimates that long-term $PM_{2.5}$ exposure accounts for roughly 7-10% of premature mortality across multiple Polish cities, despite gradual improvements in air quality. These studies collectively motivate treating ambient $PM_{2.5}$ as a central indicator within smart-city monitoring and infrastructure analytics, rather than as a narrow air-pollution statistic.

From a smart-city perspective, the relevance of $PM_{2.5}$ lies not only in its health impacts, but also in its role as a continuously monitored, regulatory-grade signal that can be integrated with operational urban datasets. Unlike specialized exposure models or dense sensor deployments, regulatory $PM_{2.5}$ observations provide a standardized, reproducible environmental baseline that is already embedded within urban governance and planning processes. When paired with routinely collected mobility indicators such as public-transit ridership, these data enable a form of city-scale observability, allowing planners and researchers to benchmark infrastructure utilization alongside exposure-relevant environmental conditions. This framing shifts the focus from attributing pollution to a single sector toward understanding how mobility intensity co-varied with ambient conditions across heterogeneous urban contexts, seasons, and scales. As such, integrating ridership and $PM_{2.5}$ data represents a practical smart-city monitoring capability, emphasizing interpretability, comparability, and scalability over model complexity.

Understanding urban $PM_{2.5}$ exposure, however, requires acknowledging that particulate concentrations are shaped by multi-sector activity and multi-scale atmospheric transport. [3] demonstrates that only a portion of population-weighted $PM_{2.5}$ originates from within city boundaries, with regional sources frequently contributing a majority share. They also show that the dominant contributing sectors vary substantially across cities, with residential, industrial, and energy-sector emissions often exceeding those from transportation. At the same time, epidemiologic studies, including [4] in the Canadian context, continue to find consistent positive associations between long-term $PM_{2.5}$ exposure and mortality even at relatively low concentration levels. Together, this evidence base provides strong justification for monitoring $PM_{2.5}$ in relation to urban activity patterns, while also recognizing the importance of structural context and seasonal conditions.

A growing body of literature has explored how transportation systems interact with particulate exposure across a range of spatial and analytical scales. At the city-ambient level, [5] reports a negative association between bus ridership and $PM_{10}$ in the Seoul Capital Region after adjusting for socioeconomic and spatial-dependence effects, suggesting that mode shift toward transit may reduce particulate pollution under some conditions. On the microenvironment scale, several studies have shown that commuter exposure varies across modes and trip segments. [6] and [7] document substantial differences between bus, subway, walking, and car travel, while [8] identifies metal-rich particulate mixtures in underground rail systems that differ from typical urban ambient $PM_{2.5}$ . Other work has focused on specific interventions or policy shocks. For example, [9] uses dispersion modeling to estimate the effects of BRT fleet electrification on roadside $PM_{2.5}$ in Bogota, while [10] analyzes London lockdown conditions as a natural experiment, finding heterogeneous pollutant responses despite large reductions in road traffic. Across these diverse approaches, three themes emerge.

First, $PM_{2.5}$ exposure remains a salient and policy-relevant urban risk indicator, including in cities without extreme pollution levels. Second, mobility systems do form part of the $PM_{2.5}$ exposure landscape, though their influence depends on baseline emissions, meteorology, and surrounding urban structure. Third, there remains a need for monitoring approaches that are simple, reproducible, and comparable across metropolitan contexts, making it possible to relate mobility intensity to regulatory $PM_{2.5}$ observations without requiring specialized modeling infrastructure or dense sensor deployments.

The present study is positioned in that space. We develop a comparative seasonal snapshot [11] analysis linking public-transit ridership and ambient $PM_{2.5}$ across four U.S. metropolitan regions, namely—New York City, Chicago, Las Vegas, and Phoenix, using agency-reported ridership data and EPA Air Quality System measurements for March and October 2024. The focus is descriptive rather than causal, and our goal is to evaluate how the apparent relationship between ridership and $PM_{2.5}$ varies when mobility is expressed in absolute versus per-capita terms, and when structural differences across cities and seasonal conditions are accounted for statistically. In doing so, we frame ridership and air-quality integration as a practical smart-city monitoring capability, supporting scale-aware interpretation of mobility and environment linkages and informing urban resilience and public-health-aware infrastructure planning.

Accordingly, this study is framed as a descriptive, scale-aware smart-city monitoring analysis that emphasizes reproducibility and cross-city comparability over causal attribution.

## II. Data Processing and Integration

This section combines regulatory air-quality observations with agency-reported public-transit data to build a harmonized, city-level panel for March and October 2024. The analysis focuses on four U.S. metropolitan regions—New York City, Chicago, Phoenix, and Las Vegas, that differ substantially in both climate and transit-use patterns. To ensure that these cities can be compared fairly, the datasets are processed using consistent temporal aggregation and population-normalization procedures.

### A. Air-Quality Data (EPA AQS)

Monthly ambient $PM_{2.5}$ concentrations were derived from the U.S. Environmental Protection Agency (EPA) Air Quality System (AQS) Daily Summary dataset. For each metropolitan region, monitors located within the constituent counties were identified, and daily arithmetic-mean $PM_{2.5}$ values were extracted for the study periods. County-level daily means were calculated by averaging across all active monitors to produce a single daily value per county. These daily observations were then averaged over the calendar month to obtain a monthly mean concentration for each city–month pair. This methodology in turn mirrors prior city-level benchmarking approaches, ensuring consistency between the descriptive and regression components of the study.

### B. Transit Ridership Data

Transit activity was measured in this study using agency-reported passenger boardings, with aggregation methods tailored to the temporal resolution of each system:

- **New York City:** In NYC, ridership data were retrieved from the Metropolitan Transportation Authority (MTA) daily subway and bus ridership statistics. Daily boardings were summed across all modes to generate a unified system-wide time series, which was subsequently aggregated to the monthly level to ensure temporal alignment with ambient air-quality data.

- **Chicago and Phoenix (Valley Metro):** For these two cities, publicly available agency reporting provides ridership at a monthly resolution. Consequently, total monthly boardings were utilized as the primary metric without further temporal interpolation.

- **Las Vegas:** In Las Vegas, ridership data were obtained from the Regional Transportation Commission of Southern Nevada (RTC) of Southern Nevada open data portal. This dataset reports passenger boardings at a granular, stop-level resolution for each month. These subtotals were summed across the entire network to yield total monthly ridership for the March and October 2024 study periods.

### C. Population Normalization

To facilitate comparative analysis across metropolitan areas of differing scale, both ridership and $PM_{2.5}$ were expressed on a per-capita basis. This normalization utilizes 2024 population estimates to account for the order-of-magnitude disparities in urban size between the study regions. Specifically, Chicago normalization used the Cook County population of 5,183,000, and Phoenix was scaled by the Maricopa County, having a population of 4,673,000. Per-capita metrics for New York City and Las Vegas were derived using total metropolitan population estimates.

Two indicators were constructed:

- **Ridership per capita:** monthly boardings divided by metropolitan population.

- **$PM_{2.5}$ per capita:** monthly mean concentration divided by metropolitan population

Although $PM_{2.5}$ per capita is not an exposure metric in a toxicological or epidemiological sense, it is used here strictly as a scale-adjusted comparative indicator. Dividing ambient concentration by metropolitan population does not represent individual inhalation risk or personal exposure. Rather, it provides a standardized lens for benchmarking environmental intensity relative to urban scale.

For example, two metropolitan regions with similar monthly mean ambient $PM_{2.5}$ concentrations but substantially different population magnitudes may reflect different system-level environmental burdens when interpreted alongside mobility activity levels. In this sense, population normalization facilitates cross-city comparability and scale-aware monitoring rather than exposure assessment.

### D. Study Period and Scope

The study scope is restricted to March and October 2024 to enable a focused seasonal comparison across the four metropolitan regions. These months were selected as representative snapshots of distinct atmospheric conditions: March reflects late-winter to early-spring conditions, while October corresponds to autumn meteorological patterns. To maintain a consistent temporal baseline and ensure uniform seasonal benchmarking, data from 2025 were excluded from the primary analysis because comparable observations were not available for all regions. This restriction minimizes inter-annual variability and allows a clearer assessment of how recurring seasonal shifts in meteorology and transit activity interact within a single annual cycle.

## III. Methodology

This study adopts a descriptive, cross-city monitoring framework designed to characterize the relationship between public-transit activity and ambient fine particulate matter ($PM_{2.5}$) under varying seasonal and structural conditions. This methodology section of the paper emphasizes transparency and reproducibility, using openly available datasets and straightforward aggregation procedures, rather than detailed emissions modeling or causal inference designs. This framework is intended to support practical, scale-aware environmental monitoring and infrastructure analytics within the smart-city context.

### A. Study Design

The analysis is structured as a comparative seasonal snapshot across four U.S. metropolitan regions: New York City, Chicago, Phoenix, and Las Vegas. To capture contrasting meteorological regimes while maintaining a rigorous temporal baseline, two representative months, March 2024 and October 2024, were selected for the study period. For each metropolitan region and month, a paired observation was constructed comprising the following variables:

- **Aggregate Transit Activity**: Total monthly system-wide public-transit ridership.
- **Environmental Baseline**: Monthly mean ambient $PM_{2.5}$ concentration derived from regulatory monitor data.
- **Normalized Metrics**: Per-capita equivalents of both transit activity and pollutant concentration, computed to account for metropolitan scale.

This multi-dimensional design facilitates a direct comparison of both absolute magnitudes and population-normalized indicators across disparate urban structures and seasonal conditions. By pairing these metrics, the framework allows for a nuanced evaluation of how mobility intensity interacts with air quality independent of city size.

### B. Mobility and Air-Quality Metrics

To evaluate the relationship between urban transit and environmental outcomes, two primary mobility variables were utilized:

- **Total Monthly Ridership:** This metric represents the aggregate sum of all passenger boardings across the metropolitan system, serving as a proxy for the total volume of transit activity.
- **Ridership Per Capita:** This variable reflects transit intensity relative to the metropolitan population size, allowing for comparisons that are independent of absolute urban magnitude.

Environmental conditions were characterized using the following indicators:

- **Monthly mean ambient $PM_{2.5}$ ($\mu g/m^3$) :** It is measured in micrograms per cubic meter, and this represents the arithmetic average of regulatory monitor readings for the study period.
- **Per-capita $PM_{2.5}$ ($\mu g/m^3$ per resident):** This is utilized as a scale-adjusted indicator to compare the relative environmental burden across cities of varying sizes.

Per-capita normalization allows for interpretation, independent of metropolitan population magnitude, ensuring that comparisons reflect usage intensity rather than city size.

### C. Cross-City Comparability

The contributing transit agencies considered in this study, differ in reporting format and service structure. Therefore, all indicators were aggregated to the monthly metropolitan level. This temporal and spatial aggregation reduces dependence on localized service characteristics (e.g., stop density, route layouts) and highlights system-level relationships between transit usage and air-quality conditions. Furthermore, the EPA AQS framework provides a rigorous regulatory baseline designed to reflect broad population-level exposure patterns. By utilizing this standardized environmental data alongside monthly transit totals, the methodology supports a consistent comparison across the four metropolitan regions, regardless of their individual infrastructure complexities.

### D. Statistical Analysis

The empirical strategy in this paper follows a structured three-step workflow to evaluate the relationship between urban mobility and environmental quality:

*1) **Descriptive comparison:*** Initial analysis summarizes the cross-city distributions of transit ridership and ambient $PM_{2.5}$. This step serves to illustrate baseline structural differences between the metropolitan regions and identify seasonal shifts between March and October 2024.

*2) **Visualization of bivariate relationships:*** Scatterplots are constructed to examine the association between ridership and $PM_{2.5}$ under both absolute and per-capita normalizations. To account for the order-of-magnitude differences in system scales, particularly when comparing New York City to smaller metros, a log-transformation of ridership is applied where appropriate.

*3) **Regression Sensitivity Analysis:*** A sequence of linear models is estimated to assess the robustness of the observed relationship as contextual controls are introduced:

*a) Pooled OLS:* Relates log-ridership directly to monthly mean $PM_{2.5}$.

*b) City Fixed-Effects Models:* Adjusts for time-invariant structural metropolitan-level baselines.

*c) Two-Way Fixed-Effects Models:* Additionally controls for month-level seasonal shocks.

Coefficient estimates and confidence intervals are reported to characterize statistical uncertainty. However, given the extremely limited sample size and degrees of freedom, these specifications are presented strictly as descriptive sensitivity checks intended to illustrate how structural and seasonal controls absorb apparent cross-sectional associations. Consistent with the study's scope, the models are used for monitoring interpretation rather than for statistical inference or causal estimation.

### *E. Analytical Assumptions and Scope*

The robustness of this methodology rests upon three primary analytical assumptions:

- **Environmental Consistency**: It is assumed that EPA AQS observations serve as a consistent proxy for ambient population-level particulate concentrations across the four metropolitan regions.
- **Data Fidelity**: Agency-reported ridership statistics are assumed to accurately reflect actual system utilization at the monthly temporal scale.
- **Seasonal Covariance**: The framework acknowledges that seasonal effects may simultaneously influence both transit demand and particulate matter dynamics.

Critically, this design does not attempt to attribute $PM_{2.5}$ variation directly to transit emissions or activity. Instead, the objective is to evaluate whether mobility intensity and exposure-relevant indicators exhibit co-movement across cities and seasons once structural metropolitan differences are statistically accounted for. This positioning identifies the framework as a tool for smart city monitoring [12] and infrastructure analytics rather than a narrow causal study.

### *F. Reproducibility Considerations*

The proposed workflow is designed to be fully replicable using public regulatory data and standard administrative aggregation procedures. By utilizing population-normalized measures, the framework ensures that monitoring capabilities can be extended to additional metropolitan regions without the requirement for specialized sensing deployments or localized emissions inventories. This approach aligns with scalable smart-city monitoring practices, offering a low-barrier methodology for integrating transit utilization and environmental benchmarks into urban resilience planning.

## IV. RESULTS AND DISCUSSION

This section examines how public transit activity aligns with ambient $PM_{2.5}$ levels using descriptive visualizations. Ridership was assembled from agency datasets for each metropolitan area: NYC MTA Daily Ridership for New York City (Subways, Buses, LIRR, Metro-North, Staten Island Railway, and Access-A-Ride), CTA bus and rail ridership for Chicago, RTC stop-level boardings for Las Vegas, and Valley Metro ridership for Phoenix. To keep the measure focused on public transit rather than roadway use, MTA Bridges and Tunnels traffic counts were excluded. Because these sources report ridership at different temporal and spatial resolutions (e.g., daily system totals versus stop-level boardings), all series were aggregated to monthly system totals for a consistent cross-city comparison. $PM_{2.5}$ was evaluated as ambient concentration ($\mu g/m^3$).

Fig. 1 illustrates total monthly transit ridership for March and October 2024. Ridership in New York City (NYC) exceeds the other systems by one to two orders of magnitude, reaching hundreds of millions of rides per month, consistent with the scale and mode share of the MTA network. NYC ridership is higher in October than in March, which may reflect seasonal shifts in travel demand. Chicago forms a distinct second tier in the tens of millions, while Las Vegas and Phoenix record the lowest totals (low millions). Because NYC largely determines the y-axis scale, variation among the smaller systems is visually compressed. Consequently, absolute ridership primarily reflects system size rather than comparable transit intensity. To account for these disparities, Fig. 4. use per-capita ridership when examining associations between mobility activity and monthly $PM_{2.5}$ concentrations.

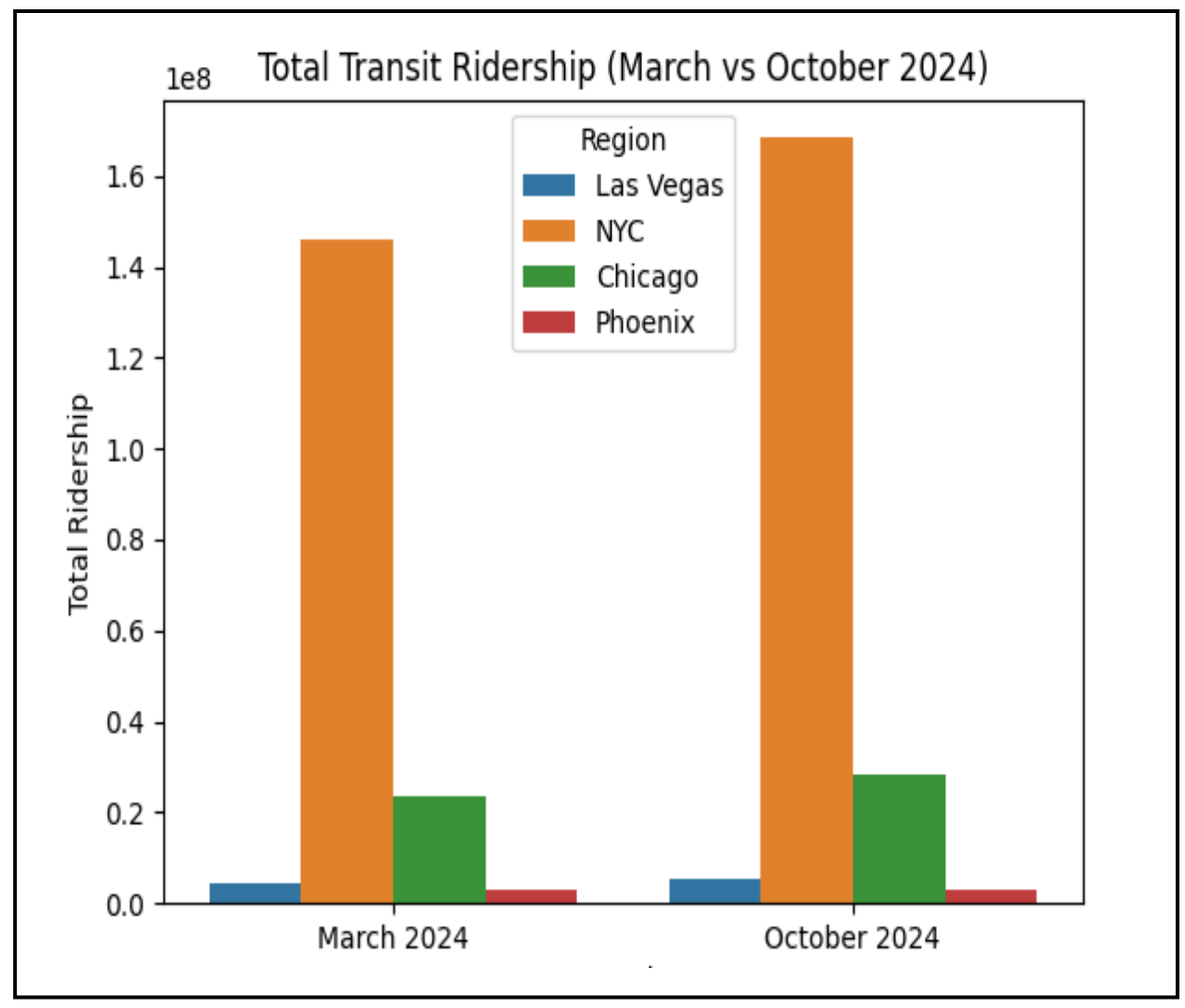

Fig. 1. Total Transit Ridership (March vs. October 2024)

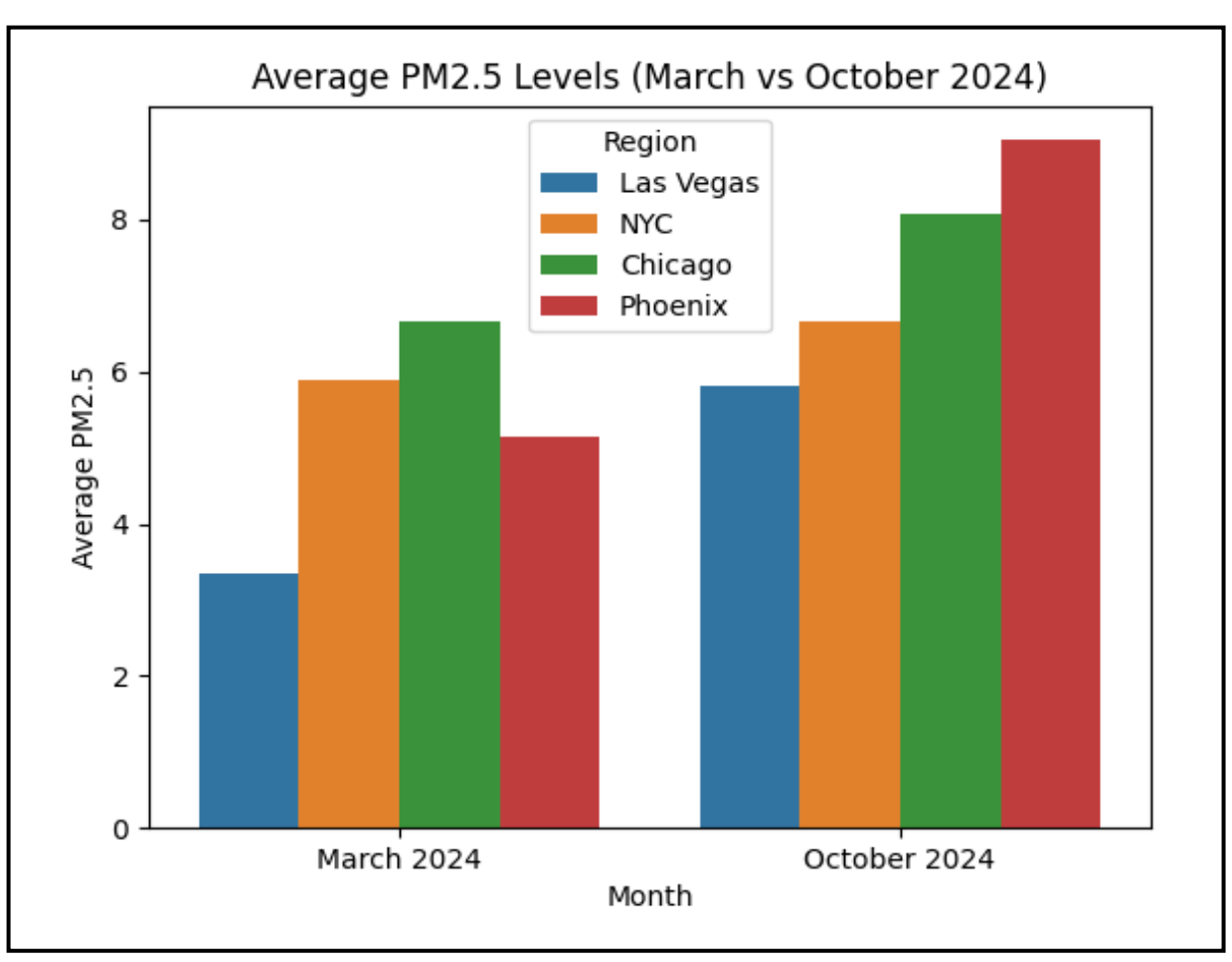

Fig. 2. Average $PM_{2.5}$ Levels (March vs. October)

Fig. 2 illustrates the monthly mean ambient $PM_{2.5}$ concentrations for March and October 2024. All four regions exhibit a consistent seasonal increase in October. While New York City and Chicago show moderate rises, the Southwestern cities like Las Vegas and Phoenix demonstrate the most pronounced seasonal shifts, with Phoenix recording the highest absolute $PM_{2.5}$ levels in October. These autumn elevations are likely associated with increased atmospheric stability and temperature inversions, which inhibit vertical mixing and thereby limit pollutant dispersion. Such effects tend to be intensified in arid climates such as Phoenix and Las Vegas, where dust contributions and residual wildfire smoke may further elevate fine-particle loads. In contrast, higher precipitation frequency and stronger ventilation in New York and Chicago may help moderate these seasonal extremes. The relative air-quality ranking also shifts between snapshots. Las Vegas records the lowest $PM_{2.5}$ in March, whereas Phoenix and Chicago emerge as the highest-concentration regions in October. Because $PM_{2.5}$ (μg/m³) reflects the combined influence of local emissions, boundary-layer dynamics, and regional transport, these fluctuations cannot be attributed to transit activity alone. This observed variability therefore motivates the subsequent pairing of ambient means with per-capita ridership indicators (Fig 4), enabling an assessment of whether mobility intensity corresponds with air-quality patterns across distinct urban environments.

Fig. 3 presents a scatterplot of monthly mean ambient $PM_{2.5}$ concentrations against total monthly public-transit ridership for the four study regions in March and October 2024. Each data point represents a city-month observation, with ridership values log-transformed ($log_{10}$) to account for the order-of-magnitude disparities in system scale. A pooled linear regression line is superimposed to summarize the aggregate cross-sectional association. The visualization indicates a weak positive correlation: higher-ridership systems, such as New York City and Chicago, generally coincide with somewhat higher ambient $PM_{2.5}$ baselines relative to the smaller systems of Phoenix and Las Vegas.

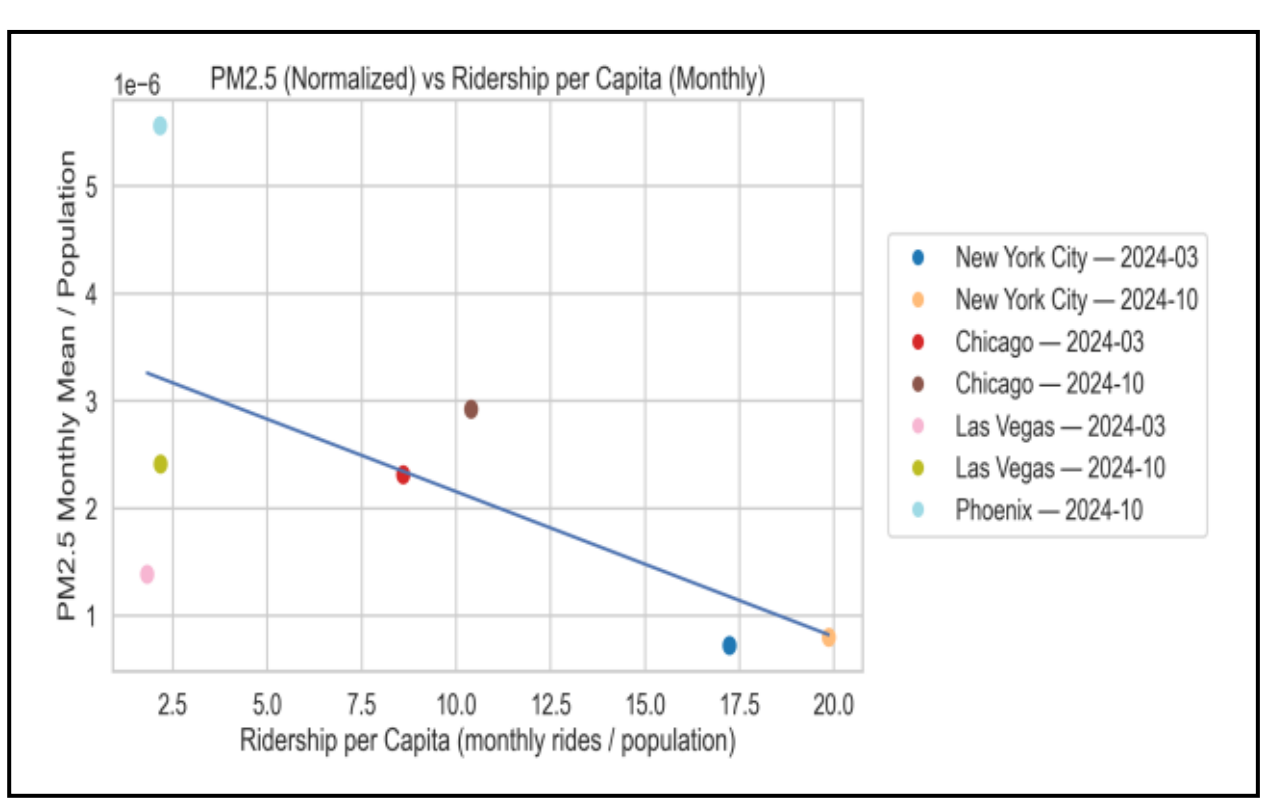

Fig. 3. $PM_{2.5}$vs log(ridership) with regression line

Accordingly, this figure is interpreted as a descriptive visualization rather than a causal estimate. The subsequent fixed-effects regression analysis (Table I) provides a descriptive sensitivity check assessing whether the weak pooled trend persists once city-specific baselines and common seasonal variation are accounted for. As detailed in later sections, the apparent association attenuates under these specifications, indicating that structural urban differences and seasonal factors explain most of the variance observed in the cross-sectional data.

However, the substantial dispersion across cities suggests that this association is shaped primarily by underlying contextual factors rather than transit activity alone. For example, although Phoenix and Las Vegas occupy the lower end of the ridership distribution, Phoenix records the highest observed $PM_{2.5}$ concentration in October (exceeding 9 μg/ $m^3$ ), consistent with the seasonal meteorological conditions discussed in Fig. 2. Conversely, New York City exhibits the largest ridership volume (on the order of $10^8$ rides per month) yet maintains moderate $PM_{2.5}$ levels relative to Chicago. These contrasts highlight the dominant influence of background meteorology, regional pollutant transport, and baseline emissions characteristics in shaping urban air quality at the monthly mean scale.

TABLE I. MODEL SUMMARY FOR TWO-WAY FIXED-EFFECTS REGRESSION OF MONTHLY MEAN $PM_{2.5}$

| **Dependent Variable** | **$PM_{2.5}$ monthly mean** |
|---|---|
| Model Type | OLS |
| Estimation Method | Least Squares |
| Number of Observations | 7 |
| Degrees of Freedom − Residuals | 1 |
| Degrees of Freedom − Model | 5 |
| Error Covariance Assumption | Non robust |
| R-squared | 0.983 |
| Adjusted R-squared | 0.899 |

| | coef | std err | [0.025 | 0.975] |
|---|---|---|---|---|
| log_rides | 63.5778 | 52.296 | -600.91 | 728.066 |

[a] Given the extremely small sample size, coefficients are reported for descriptive sensitivity only.

Table I presents a two-way fixed-effects (TWFE) sensitivity specification intended to illustrate how apparent cross-sectional associations between transit ridership and ambient $PM_{2.5}$ are absorbed once structural city differences and common seasonal factors are introduced. With both city and month indicators included, the estimated ridership coefficient becomes highly imprecise and unstable, reflecting that most observed variation in monthly mean $PM_{2.5}$ at the city-month scale is accounted for by persistent metropolitan characteristics and shared seasonal conditions rather than short-term ridership fluctuations.

Given the extremely limited panel size and degrees of freedom, coefficient magnitudes are not interpreted substantively. Instead, the TWFE specification serves as a descriptive benchmarking exercise, demonstrating how contextual controls reshape apparent mobility–environment relationships in small cross-city monitoring panels.

Fig. 4 examines the relationship between population-normalized ambient $PM_{2.5}$concentrations and monthly public-transit ridership per capita for the four study regions in March and October 2024. Ridership per capita is defined as total monthly system boardings divided by the metropolitan population. A pooled linear regression line is superimposed to summarize the aggregate trend across the normalized observations. In contrast to the raw-ridership specification in Fig. 3, the per-capita normalization yields a modest negative slope: cities and months with higher per-capita transit usage tend to coincide with lower population-normalized $PM_{2.5}$levels in this sample.

Some dispersion remains among city–month observations, and the limited sample size motivates a descriptive rather than causal interpretation. Nevertheless, the figure highlights that per-capita scaling alters the apparent relationship between mobility intensity and urban air-quality exposure. The reversal in the regression slope underscores the importance of accounting for metropolitan scale when evaluating associations between transit utilization and environmental indicators.

Taken together, the patterns in Fig. 3 and Fig. 4 suggest that the way mobility is scaled plays a key role in how public ridership vs. air-quality relationships appear at the city level. When ridership totals are compared directly across metropolitan areas, the results tend to mirror overall city size and activity levels. Once these totals are normalized by population, however, the comparison shifts toward how intensively residents rely on transit rather than how large the system is.

This distinction matters because many environmental indicators, including $PM_{2.5}$ are closely tied to the broader urban fabric, from land-use intensity to regional emissions baselines. Per-capita mobility metrics provide a way to separate those background effects from differences in how people actually move around within a city. In that sense, the per-capita framing offers a more scale-independent lens for interpreting transit usage alongside exposure metrics. While the sample here is intentionally small, the contrast between the two normalizations highlights the value of carefully selecting the unit of comparison in smart city monitoring and planning.

Table II reports pooled and fixed-effects linear regression models relating log-transformed monthly public-transit ridership to monthly mean ambient $\mathrm{PM}_{2.5}$ concentrations across the four study regions. Model (1) estimates a pooled ordinary least-squares specification without city or month controls. Model (2) introduces city fixed effects to account for structural differences across metropolitan systems. Model (3) further includes month fixed effects (FE), capturing seasonal and meteorological heterogeneity common across cities.

Across specifications, the apparent association between ridership and $\mathrm{PM}_{2.5}$ attenuates substantially once structural and seasonal controls are introduced. This pattern illustrates those cross-sectional differences in ambient $\mathrm{PM}_{2.5}$ at the monthly scale dominated by metropolitan baseline characteristics and seasonal dynamics rather than short-term fluctuations in transit activity.

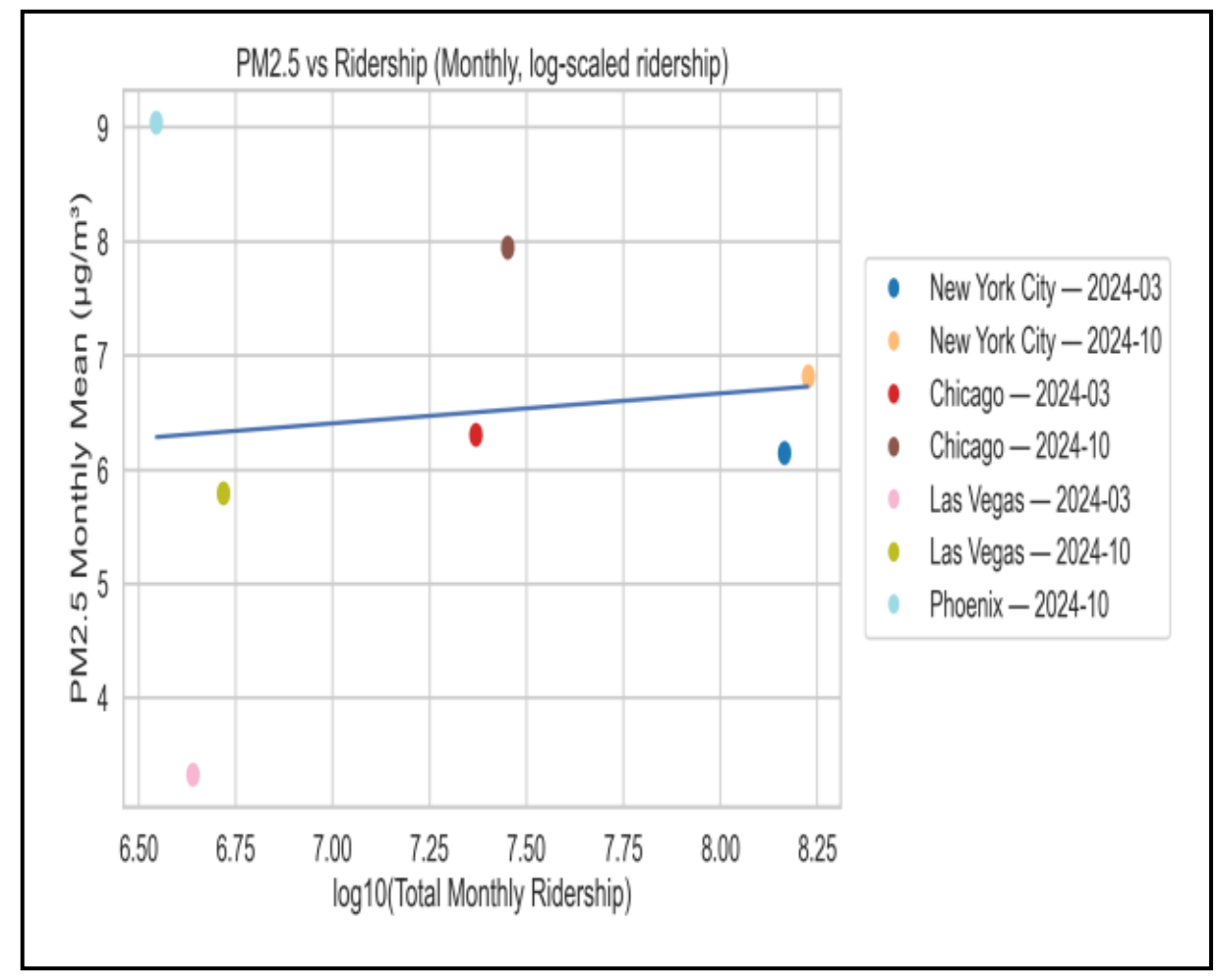

Fig. 4. $PM_{2.5}$ vs ridership per capita

TABLE II. POOLED AND FIXED-EFFECTS REGRESSION MODELS RELATING TRANSIT RIDERSHIP TO MONTHLY MEAN $PM_{2.5}$

| Dependent Variable: Monthly Mean PM2.5 (µg/m³) | | | |
|---|---|---|---|
| | **(1) Pooled OLS** | **(2) City Fixed Effects** | **(3) City + Month FE** |
| log10(Ridership) | 0.263<br>(1.135) | 22.110<br>(5.661) | 63.578<br>(52.296) |
| Constant | 4.567<br>(8.317) | -156.702<br>(41.953) | -462.424<br>(385.579) |
| City Fixed Effects | No | Yes | Yes |
| Month Fixed Effects | No | No | Yes |
| Observations | 7 | 7 | 7 |
| R-squared | 0.01 | 0.97 | 0.98 |
| Adj. R-squared | -0.19 | 0.92 | 0.90 |
| Notes: Standard errors in parentheses. Model (1) estimates a pooled OLS specification without city or month controls. Model (2) includes city fixed effects. Model (3) includes both city and month fixed effects. PM2.5 values reflect EPA AQS monthly means; ridership reflects agency monthly totals. | | | |

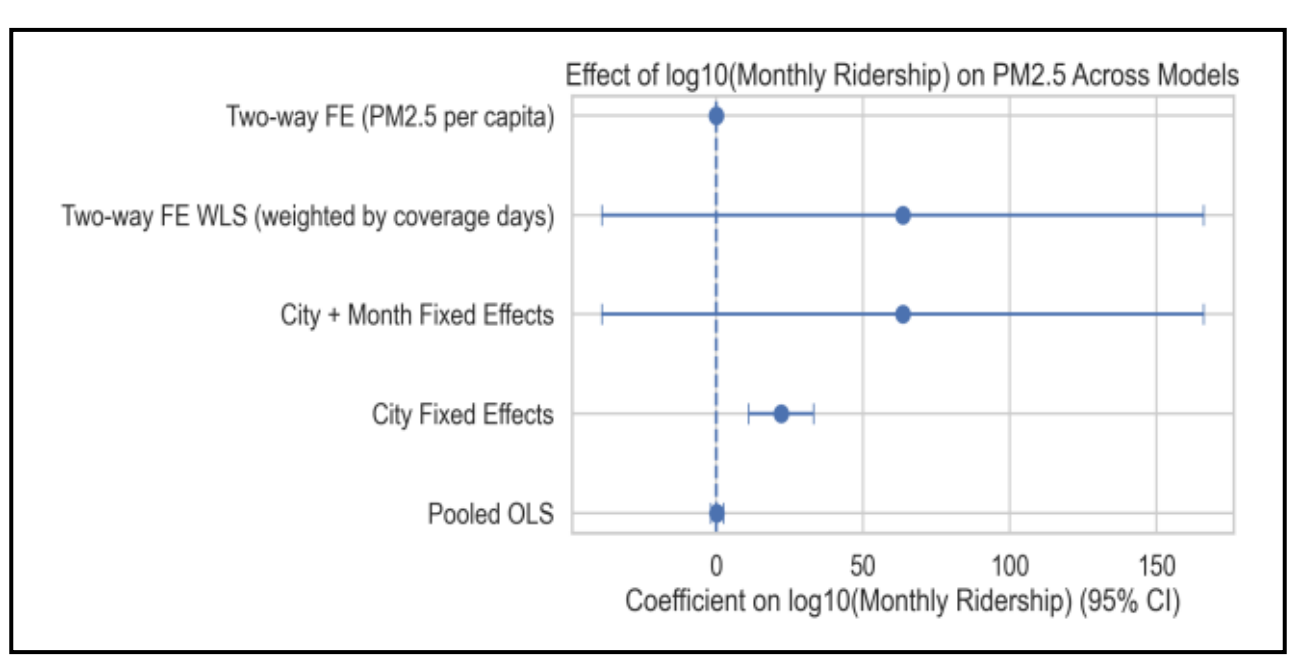

Fig. 5. Effect of log10 (Monthly Ridership) on $\mathrm{PM}_{2.5}$ Across Model

From a monitoring perspective, the primary value of these models lies not in coefficient magnitude but in demonstrating how normalization strategies and contextual controls alter interpretation. The regression workflow therefore functions as a descriptive sensitivity analysis that reinforces the importance of scale-aware benchmarking when integrating operational transit metrics with regulatory environmental indicators. Given the limited sample size, the results are presented for illustrative purposes only and are not intended to support causal or inferential claims.

Fig. 5. summarizes the estimated coefficient on $\log_{10}$(Monthly Ridership) across three core regression specifications: (i) a pooled ordinary least-squares (OLS) model, (ii) a city fixed-effects model, and (iii) a two-way fixed-effects (TWFE) model that controls for both city-level baselines and month-specific seasonal variation. Two extensions of the TWFE model are also included: one weighted by the number of valid $PM_{2.5}$ reporting days and another using per-capita $PM_{2.5}$ as the dependent variable. Horizontal bars denote 95% confidence intervals.

The pooled OLS and city fixed-effects models suggest a positive association between ridership and $PM_{2.5}$. However,

this apparent association attenuates under the TWFE specifications once structural and seasonal controls are introduced. Once concurrent city-specific baselines and common seasonal shocks are controlled, the point estimates shrink toward zero and become statistically indistinguishable from no effect. This indicates that the raw correlation between transit volume and air quality is likely shaped by background heterogeneity, such as metropolitan scale, land-use intensity, and meteorology rather than transit activity itself.

Taken together, these results reinforce the broader conclusion of this study: variation in ambient $PM_{2.5}$ at the city-month scale appears to be dominated by regional and seasonal forces, not short-term differences in transit ridership. Accordingly, the regression models are interpreted as descriptive sensitivity analyses rather than causal estimates, highlighting the importance of structural and temporal controls in urban environmental analytics.

From a monitoring perspective, the primary value of these regression specifications lies not in coefficient magnitude but in demonstrating how normalization strategies and contextual controls reshape apparent mobility–environment relationships. This reinforces the importance of scale-aware benchmarking when integrating operational transit metrics with regulatory environmental indicators.

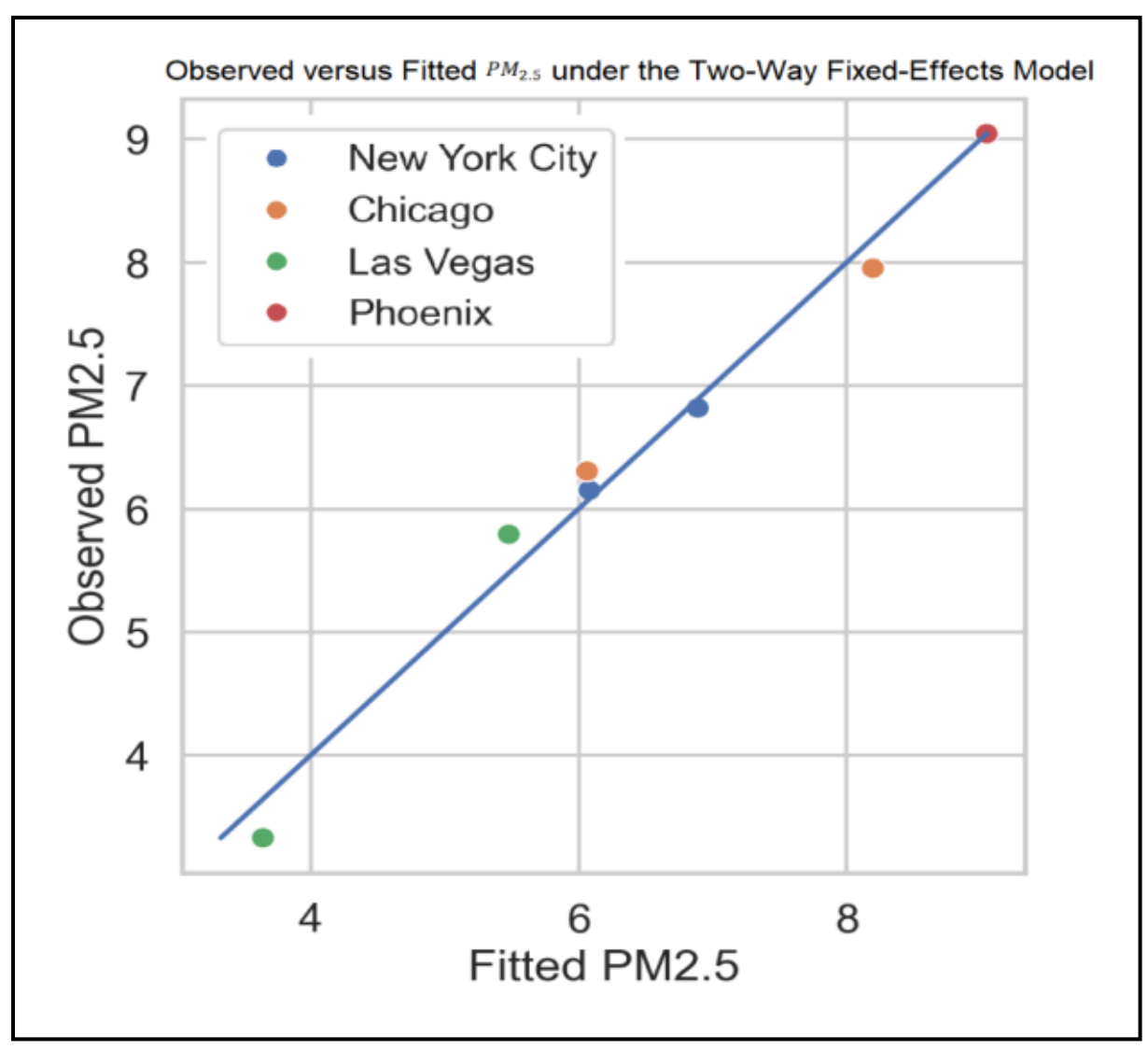

Fig. 6. Observed versus Fitted $PM_{2.5}$under the Two-Way Fixed-Effects Model

Fig. 6. compares the observed monthly mean $PM_{2.5}$ concentrations with the corresponding fitted values from the two-way fixed-effects (TWFE) regression model that includes both city and month indicators. Each point represents a city-month observation, and the 45-degree reference line indicates perfect fit. The close alignment of the points with the reference line reflects the strong explanatory power of the fixed-effects structure, which captures persistent city-level differences in baseline air quality as well as seasonal variation common across cities. This pattern is consistent with the regression results showing that most of the variance in $PM_{2.5}$ is explained by structural and temporal heterogeneity rather than by changes in transit ridership. Accordingly, the TWFE framework primarily acts to absorb background environmental and meteorological conditions while the ridership coefficient remains small and statistically indistinguishable from zero.

Taken together, these results show that the apparent relationship between public-transit ridership and ambient $PM_{2.5}$ depends strongly on how mobility is measured and contextualized. In raw aggregate form, ridership mainly reflects city scale and baseline activity rather than true transit intensity. Once population normalization and fixed-effects controls are applied, short-term ridership variation explains little of the remaining variation in monthly $PM_{2.5}$ , which is instead dominated by structural city characteristics and seasonal conditions.

For smart-city analytics, this highlights the importance of scale-aware, context-sensitive integration of mobility and environmental data. Simple linkage of operational ridership totals to air-quality observations, without appropriate normalization, risks misleading interpretation. By contrast, the population-normalized and fixed-effects framework used here offers a lightweight and reproducible way to benchmark cities while preserving interpretability, supporting evidence-based planning and resilient infrastructure management.

## V. Limitations

This study is designed as a descriptive monitoring analysis rather than a causal investigation, and several limitations apply. The dataset includes only two seasonal snapshots (March and October 2024), so it does not represent full annual variability. As a result, observed patterns may be sensitive to short-term meteorological or structural conditions specific to the selected months. The resulting panel is intentionally small, and regression specifications are presented strictly as illustrative sensitivity checks rather than for statistical inference. In particular, the two-way fixed-effects specification leaves only one residual degree of freedom, reinforcing that coefficient estimates are not suitable for inferential interpretation.

Monthly mean ambient $PM_{2.5}$ values reflect broad population-level ambient conditions but do not identify emission sources or capture localized microenvironments. While a per-capita normalization of $PM_{2.5}$ is used to facilitate cross-city scale comparison, this metric is not an exposure indicator in a toxicological sense and should be interpreted strictly as a scale-adjusted benchmarking tool.

Transit ridership is treated as an aggregate system-level indicator, meaning residual confounding from economic activity, freight movement, industrial emissions, wildfire smoke, and meteorological variation may remain. Future extensions incorporating explicit meteorological covariates and expanded temporal coverage would strengthen robustness.

Finally, the analysis is limited to four U.S. metropolitan regions, and results may not generalize universally.

Even with these constraints, the framework shows that simple, reproducible integration of mobility and regulatory air-quality data can support scale-aware urban monitoring and benchmarking.

## VI. Conclusions and Future Work

This paper develops a simple, reproducible framework for integrating public-transit ridership with regulatory ambient $PM_{2.5}$ measurements as part of a smart-city monitoring capability. Using March and October 2024 as seasonal benchmarks across four U.S. metropolitan regions, the results show that the apparent relationship between ridership and air quality depends strongly on population scaling and structural controls. In aggregate form, higher ridership mainly reflects metropolitan scale. After population normalization and fixed-effects adjustment, short-term ridership variation explains little of the remaining variation in monthly $PM_{2.5}$, which is instead driven largely by structural and seasonal factors.

These findings emphasize the need for scale-aware, context-sensitive interpretation when combining mobility and environmental indicators in smart-city analytics. The proposed framework shows that openly available data can be integrated transparently to support benchmarking and public health-aware planning without requiring specialized sensing infrastructure.

From an operational standpoint, municipal transportation agencies or sustainability offices could implement this framework as a lightweight monitoring dashboard that integrates routinely reported ridership totals with regulatory AQS air-quality data. Such a system could support seasonal benchmarking, infrastructure planning, and cross-city comparison without requiring new sensing deployments. For example, agencies could use this approach to track whether changes in transit intensity coincide with broader environmental trends over time, helping to contextualize mobility investments within public-health and resilience planning objectives.

The primary contribution of this work is the demonstration of a scale-aware, reproducible smart-city monitoring framework that integrates public-transit utilization with regulatory ambient $PM_{2.5}$ data, rather than the estimation of causal impacts of transit on air pollution.

Future work will extend the framework to additional cities and years and incorporate basic meteorological covariates, such as temperature, wind speed, boundary-layer stability, precipitation, and wildfire smoke indicators to further disentangle seasonal atmospheric dynamics from structural urban effects. Additional enrichment through city-type descriptors (e.g., climate zone, land-use intensity, transit mode share) would allow the framework to be incrementally expanded while preserving its monitoring orientation. Such extensions would enhance robustness while maintaining the reproducible, scalable character of the proposed approach.